\documentclass[journal]{IEEEtran}

\usepackage{cite}

\ifCLASSINFOpdf
  \usepackage[pdftex]{graphicx}
  \usepackage{epstopdf}
\else
\fi
\usepackage{amsmath}
\usepackage{amsfonts}

\ifCLASSOPTIONcompsoc
  \usepackage[caption=false,font=normalsize,labelfont=sf,textfont=sf]{subfig}
\else
  \usepackage[caption=false,font=footnotesize]{subfig}
\fi

\usepackage{color,soul}
\begin{document}
%
\title{Limitations of the intrinsic cut-off frequency to correctly quantify the speed of nanoscale transistors}
\author{Zhen Zhan, Enrique Colom\'es, Xavier Oriols,~\IEEEmembership{Member, IEEE}

\thanks{The authors are with the Departament d'Enginyeria Electr\`onica, Universitat Aut\`onoma de Barcelona, Bellaterra, 08193, Spain (e-mail: xavier.oriols@uab.cat).
This work has been partially supported by the Fondo Europeo de Desarrollo Regional (FEDER), the ``Ministerio de Ciencia e Innovaci\'{o}n'' through the Spanish Project TEC2015-67462-C2-1-R,  the Generalitat de Catalunya (2014 SGR-384) and the European Union's Horizon 2020 research and innovation programme under grant agreement No 696656. Z. Zhan acknowledges financial support from the China Scholarship Council.
}}

\maketitle

\begin{abstract}
The definition of the intrinsic cut-off frequency ($f_T$) based on the current gain equals to one (0 dB) is critically analyzed. A condition for the validity of the quasi-static estimation of  $f_T$ is established in terms of the temporal variations of the electric charge and electric flux on the drain, source and gate terminals. Due to the displacement current, an electron traversing the channel length generates a current pulse of finite temporal width. For electron devices where the intrinsic delay time of the current after a transient perturbation is comparable to such width, the displacement currents cannot be neglected and the quasi-static approximation becomes inaccurate. We provide numerical results for some ballistic transistors where the estimation of $f_T$ under the quasi-static approximation can be one order of magnitude larger than predictions obtained from a time-dependent numerical simulations of the intrinsic delay time (including particle and displacement currents). In other ballistic transistors, we show that the gate current phasor can be smaller than the drain one at all frequencies, giving no finite value for $f_T$. 
\end{abstract}

\begin{IEEEkeywords}
Cut-off frequency, THz, displacement current, nano transistor, time-dependent simulation.
\end{IEEEkeywords}

\section{Introduction}

\IEEEPARstart{T}{he} development of faster electron devices for digital and analog applications is a constant demand in the electronics industry \cite{Schwierz2010,Fiori2014}.The scientific community tries to quantify how fast the field effect transistors (FETs) work through some figures of merit (FoMs). Unfortunately, there is no such a unique FoM that unequivocally quantifies the speed. Some definitions are linked to a particular circuit or application, others to the intrinsic device itself. Some FoMs are redefined to make them more easily accessible from simulations, or from measurements. Usually, the FoM in digital FET applications are related with times, while in analog ones are commonly described with frequencies. 

In digital FET applications, for example, an important FoM is the intrinsic delay time $\tau_d$. The idea of this FoM is to quantify the time needed for an output signal to respond to an input signal\cite{Taur}. Many times, a simpler quasi-static definition of the intrinsic delay time, $\tau_d^{QS} \approx C \cdot V_{gs}/I_{ds}$, is preferred because it is easily accessible from DC (time-independent) simulations \cite{Schwierz2010}. Such expression can be interpreted as the time needed to charge the next gate capacitor $C$ until the gate voltage $V_{gs}$ associated to the ON state, with a constant drain-source current $I_{ds}$.  From an experimental point of view, however, a new definition of the intrinsic delay time from a ring oscillator of $N$ (odd) CMOS inverters is used. By taking the inverse of the frequency at which the ring oscillator runs and dividing it by $N$, such intrinsic delay time can be easily obtained \cite{Shohno1978}.  

In analog applications, the cut-off frequency $f_T$ and the maximum oscillation frequency $f_{max}$ are the main FoMs. The $f_T$ is defined as the frequency at which the drain and gate currents become equal (that is a current gain of 0 dB) \cite{Laux1985,Wang2015}. Equivalently, the $f_{max}$ is the frequency at which the power gain is 0 dB \cite{Schwierz2010}. Both frequencies are easily accessible from the measurement of $S$-parameters and even their intrinsic values (when all parasitic elements of the circuit are eliminated using de-embedding techniques) are measurable. Needless to say, $f_{max}$, based on Mason's identities \cite{Mason1954}, becomes a more relevant FoM in high-frequency analog applications \cite{Fiori2014}.   

It is accepted that, although the intrinsic $f_T$ is not the relevant FoM in high frequency analog applications, it is a meritorious FoM providing useful information on the speed of FETs. In order to provide an expression of the cut-off frequency accessible from DC (time-independent) simulations,  the so-called quasi-static approximation $f_T^{QS} \approx g_m/(2 \pi C)$ is presented in the literature \cite{Venica2014,Logoteta2014,Cho2011,Zheng2016,Dong2017}. It is based on assuming that the drain current is only the DC component related to the (linear) transconductance $g_m=dI_{ds}/dV_{gs} \approx I_{ds}/V_{gs}$, while the gate current is the displacement component proportional to the capacitor $C$ and frequency. From the previous quasi-static definition of the intrinsic delay time $\tau_d^{QS}\approx C \cdot V_{gs}/I_{ds}$ in digital applications, we easily arrive to the approximation $f_T^{QS} \approx {1}/{(2 \pi \tau_d^{QS})}$ \cite{Singh2001,Rutherglen2009}. This last expression supposedly justifies why the cut-off frequency is a good FoM to quantify the intrinsic switching speed in digital applications. Alternatively, several non-quasi-static approximations are also proposed for more accurate predictions of $f_T$ \cite{Wang2015,Cho2011,Teppati2014,Cuchillo2016}. 

In this paper, we discuss if $f_T$ can be an appropriate FoM to quantify the intrinsic speed of these nanoscale FETs with dimensions of few nanometers for digital or analog TeraHertz (THz) applications. In such FETs, the electric field generated by an electron crossing the channel is not properly screened and it induces displacement current on the terminals. We will construct a condition for the validity of the quasi-static estimation of $f_T^{QS}$ and prove that $f_T$ can be a quite misleading estimator (with or without approximations) for the speed of ballistic FETs. 

We summarize here four relevant time intervals that will be used along the paper. The intrinsic delay time $\tau_d$ quantifies the temporal difference between the time when a gate voltage perturbation starts and the time when the gate, drain and source currents reach steady-state values. The $\tau_d^{QS}$ and $\tau_d^{NQS}$ are the intrinsic delay time mentioned above under the quasi-static approximation and under the zero-order non-quasi-static approximation, respectively. Due to the displacement current, one electron traversing the channel length generates a current pulse \footnote{In the case of an electron with charge $q$ and velocity $v$ moving between two infinite parallel metals separated with a distance $L$, it is well known from the Ramo-Shockley-Pellegrini theorem \cite{Oriols2013} that the square pulse current has a temporal width of $\tau_p=L/v$ and a height equals to $qv/L$. The total charge of the current pulse is $(qv/L)\times(L/v)=q$.}. The temporal width of such pulse is defined as $\tau_p$. The value of $\tau_p$ is influenced by the device geometry and the dielectric relaxation time needed for the background charge to neutralize (screen) the electric field generated by the single electron. In the text, we also define $f_T^d$, $f_T^{QS}$ and $f_T^{NQS}$ as the cut-off frequencies associated to $\tau_d$, $\tau_d^{QS}$ and $\tau_d^{NQS}$, respectively. Finally, $f_T$ is the exact definition of the cut-off frequency from the current gain equals to one. We will see in this paper that $f_T$ can differ from $f_T^d$. 

\section{Fourier Analysis of $f_T$}

In this Section, a Fourier analysis of the definition of $f_T$ from the current gain equals to one will be discussed, with special attention to the role played by the particle and displacement currents on it. This complete discussion is valid for any type of (ballistic or non-ballistic) FETs. We will also present the conditions of validity of the quasi-static approximation in terms of the temporal variations of the electric charge plus the electric flux on the drain, source and gate terminals.

\subsection{Preliminary Discussion}
\label{preli}

\begin{figure}[t!]
\centering
\subfloat[]{\includegraphics[width=0.5\columnwidth]{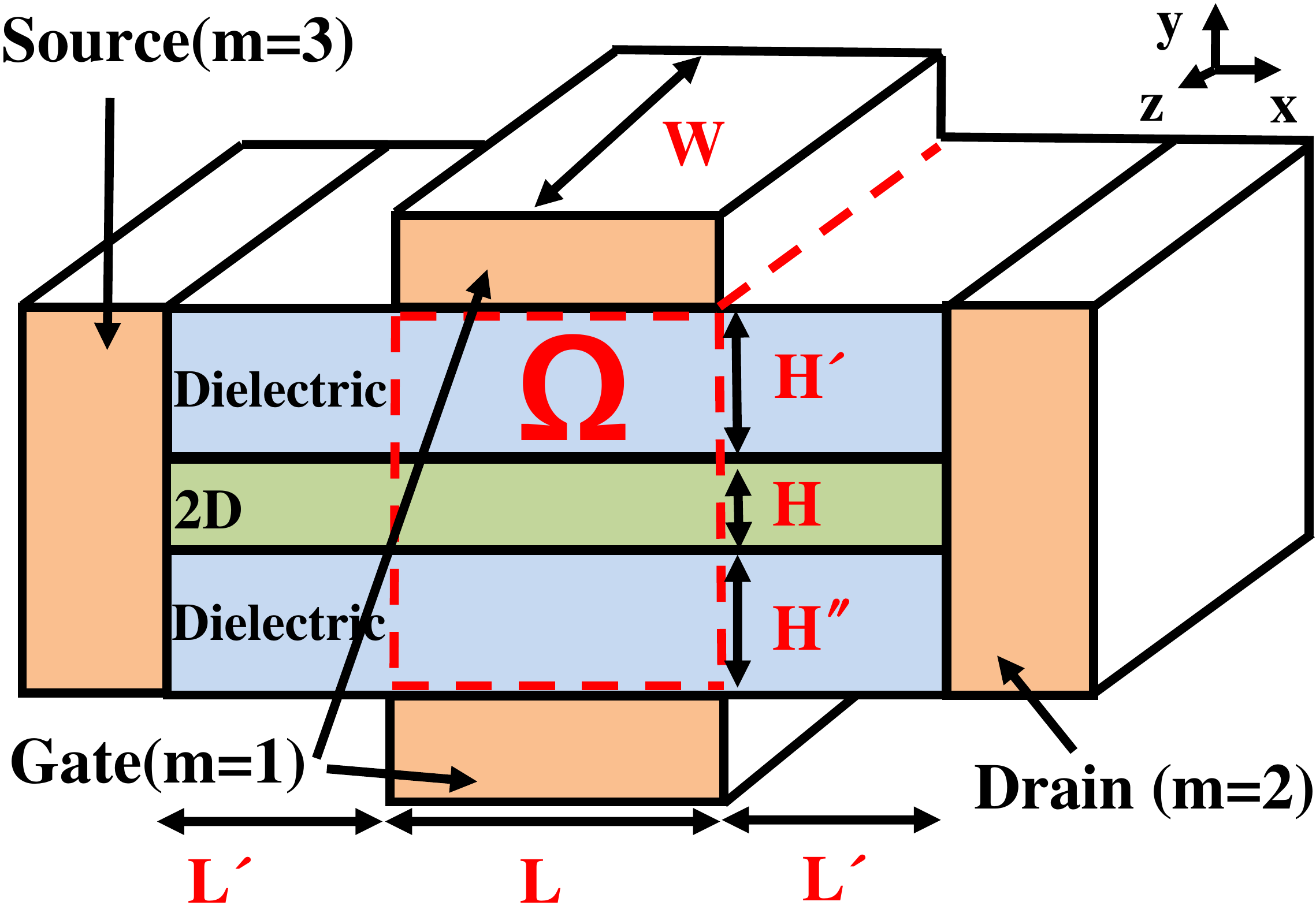}\label{cross_section}}
\subfloat[]{\includegraphics[width=0.5\columnwidth]{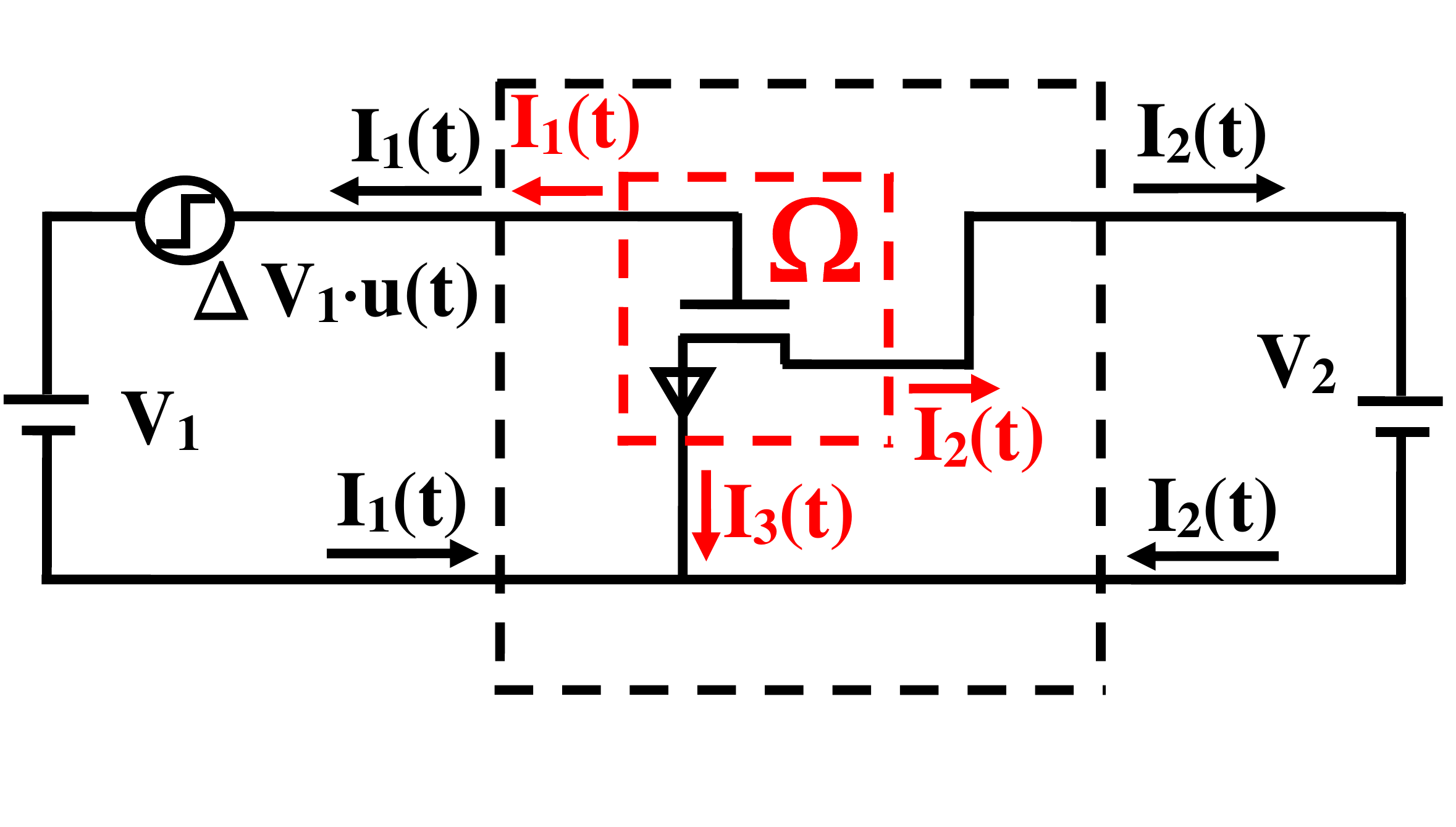}\label{cross_section}}
\caption{(a) Cross-section of the active region $\Omega=L \times (H'+H+H'') \times W$ in a dual-gate 2D FET being $L$ the gate length.  (b) A two-port network of the intrinsic device. }
\label{figure1}
\end{figure}

We consider a dual-gate FET depicted in Fig. \ref{figure1}(a) with three terminals. The three relevant total (displacement plus particle) currents, named $I_1(t)$, $I_2(t)$ and $I_3(t)$ are associated to the gate, drain and source terminals, respectively, as
\begin{equation}
I_m(t)=\int_{S_m} \epsilon(\vec{r}) \frac{\mathrm d \vec E(\vec r,t)}{\mathrm d t} \cdot \mathrm{d}\vec s+\int_{S_m} \vec J(\vec r,t) \cdot \mathrm{d}\vec s,
\label{total}
\end{equation}
being $\epsilon(\vec{r})$ the electric permittivity, $\vec E(\vec r,t)$ the electric field and $\vec J(\vec r,t)$ the particle current density. We consider $\mathrm{d}\vec s$ outwards. The three surfaces in equation (\ref{total}) construct a surface $S=S_1+S_2+S_3$ that totally enclose an arbitrary volume $\Omega$. Then, by construction, at any time $t$, the three currents satisfy 
\begin{equation}
I_1(t)+I_2(t)+I_3(t)=0.
\label{transient_current}
\end{equation}
which is just the conservation of the total current in the active region $\Omega$ due to the application of Gauss's law in $S$. 

In the evaluation of $f_T$, we are interested in a transient simulation. Initially, the three currents have steady-state values $I_m(0)$. At $t=0$, a (small-signal) voltage perturbation is applied on one of the three FET terminals. Then,  during a time interval $\tau_d$ (we will see later that this time is indeed the intrinsic delay time), the three output currents oscillate. Finally, new steady-state values $I_m(\tau_d)$ for the three currents are achieved with $I_m(t)=I_m(\tau_d)$ for $t\geq \tau_d$. For each contact $m$, we define the incremental charge during $\tau_d$ as
\begin{eqnarray}
\Delta Q_m \equiv \int_{0}^{\tau_d} (I_m(t)-I_m(\tau_d))\mathrm{d}t.
\label{Q}
\end{eqnarray}
Using (\ref{transient_current}) that ensures $I_1(t)+I_2(t)+I_3(t)-I_1(\tau_d)-I_2(\tau_d)-I_3(\tau_d)=0$ at any time, we easily get
\begin{equation}
\label{charge}
\Delta Q_1+\Delta Q_2+\Delta Q_3=0.
\end{equation} 
This equation just states that the variation of electron charge from $0$ till $\tau_d$  in the volume $\Omega$ is compensated by the variations of the electric flux during this time interval on the surface $S$. 

\subsection{The All-order Definition of $f_T$}

The usual definition of $f_T$ comes from a FET in common source configuration as plotted in Fig. \ref{figure1}(b). Following the signs of the currents assigned to the FET of Fig. \ref{figure1}(a), the currents on the gate and drain terminals of the two-port network are positive when leaving the network.  In the two-port network model of Fig. \ref{figure1}(b), the relationship between the phasor voltages $\tilde V_m(\omega) \equiv \mathcal F\{V_m(t)-V_m(0)\}$, the phasor currents $\tilde I_m(\omega) \equiv \mathcal F\{I_m(t)-I_m(0)\}$ and the $Y$-parameter matrix is
\begin{equation}
\left[\!
    \begin{array}{c}
      \tilde I_1(\omega) \\
     \tilde I_2(\omega)
    \end{array} 
\!\right]=
\left[\!
\begin{array}{cc}
      Y_{11}(\omega) \;\; Y_{12}(\omega)\\
     Y_{21}(\omega)  \;\; Y_{22}(\omega)
    \end{array}
\!\right]
\left[\!
    \begin{array}{c}
      \tilde V_1(\omega) \\
     \tilde V_2(\omega)
    \end{array} 
\!\right].
 \label{2_port}
\end{equation}
where $\omega$ is the angular frequency and $\mathcal F\{..\}$ is the Fourier transform. The frequency-dependent component $Y_{mn}(\omega)$ of the admittance matrix, due to a small signal current $I_m(t)-I_m(0)$ collected on contact $m$ when a step perturbation $V_n(t)=V_n(0)+\Delta V_n\cdot u(t)$ (with $u$ the Heaviside step function) is applied on contact $n$ and zero volts in the rest of terminals, is given by \cite{Laux1985,Hockney}
\begin{eqnarray}
\label{Yij}
Y_{mn}(\omega) \equiv \frac{\Delta I_m}{\Delta V_n}+\frac{j\omega}{\Delta V_n} \int_{0}^{\tau_d}\left(I_m(t)-I_m(\tau_d)\right)\;e^{-j \omega t}dt,
\end{eqnarray}
where $\Delta I_m=I_m(\tau_d)-I_m(0)$. In expression (\ref{Yij}), we have considered $I_m(t)=I_m(\tau_d)$ for $t\geq \tau_d$.  

The intrinsic cut-off frequency $f_T$ computed from the $Y$-parameter is the linear frequency at which the current gain magnitude drops to unity (0 dB) \cite{Laux1985} 
\begin{equation}
\label{current_gain1}
|h_{21}(\omega=2\pi f_T)| \equiv \frac {\left| Y_{21}^{All} \right|} {\left| Y_{11}^{All} \right|}\equiv \frac{\mathcal F\{I_2(t)-I_2(0)\}}{\mathcal F\{I_1(t)-I_1(0)\}}=1.
\end{equation}
The superindex ${All}$ means that all orders of the Taylor expansions of the Fourier transform in (\ref{Yij}) are taken into account (without approximations).

\subsection{The Quasi-static Definition of $f_T^{QS}$}
\label{qs}

The expression of $f_T^{QS}$ within the quasi-static approximation is obtained by computing the term $Y_{21}$ from (\ref{Yij}) without any frequency dependence, $ e^{-j\omega t}\approx 0$, as
\begin{equation}
 Y_{21}^{QS}(\omega) \approx \frac{\Delta I_2}{\Delta V_1} \equiv \frac{dI_2}{dV_1}\equiv g_m. \label{Y_{21}1}
\end{equation}
The term $Y_{11}$ is computed with a zero-order approximation, $ e^{-j\omega t}\approx 1$, from (\ref{Yij}) as
\begin{equation}
 Y_{11}^{QS}(\omega) \approx \frac{j\omega}{\Delta V_1}\int_{0}^{\tau_d} (I_1(t)-I_1(\tau_d)) \;{\rm d}t= j\omega\frac{\Delta Q_1}{\Delta V_1},   \label{Y_{11}1}
\end{equation}
where $\Delta Q_1$ is defined in (\ref{Q}). The approximation in (\ref{Y_{21}1}) is based on the assumption that the current pulse $\tau_p$ is short enough to neglect any displacement component of the drain current. Expression (\ref{Y_{11}1}) assumes that the gate current is the displacement component. As indicated in the Introduction, from (\ref{current_gain1}), using (\ref{Y_{21}1})  and (\ref{Y_{11}1}) , we get
\begin{equation}
f_{T}^{QS} =\frac{g_m}{2\pi \Delta Q_1/\Delta V_1}=\frac{g_m}{2\pi C_1}, 
\label{f_t_QS}
\end{equation}
where the term $\Delta Q_1/\Delta V_1 \equiv C_1$ is usually associated to the gate capacitor \cite{Cheng2001}. If we assume that $\Delta Q_1 \approx \Delta Q_2$, during the transient evolution 
\begin{equation}
\Delta Q_1 \approx \Delta Q_2  \equiv \int_{0}^{\tau_d} (I_2(t)-I_2(\tau_d))\mathrm{d}t \approx \Delta I_2 \tau_d^{QS},
\label{aprox}
\end{equation}
we get the condition $C_1 \Delta V_1= \Delta Q_1 \approx \Delta Q_2 \approx \Delta I_2 \tau_d^{QS}$ where $\tau_d^{QS} \approx C_1 \Delta V_1 / \Delta I_2 $ is the typical quasi-static definition of the intrinsic delay time  mentioned in the Introduction when $\Delta V_1 \approx V_{gs}$ giving $\Delta I_2 \approx I_{ds}$. Then, the definition of the (small-signal) transconductance in equation (\ref{Y_{21}1}), with expression (\ref{aprox}), can be redefined as \cite{Singh2001}
\begin{equation}
g_m \equiv \frac{dI_2}{dV_1} \approx \frac{\Delta I_2}{\Delta V_1}=\frac {\Delta I_2} {\Delta Q_1} \frac {\Delta Q_1} {\Delta V_1}=\frac {1} {\tau_d^{QS}} C_1.
\label{gm_QS}
\end{equation}
Putting (\ref{gm_QS}) into (\ref{f_t_QS}), one arrives to the final result
\begin{equation}
f_{T}^{QS} =1/(2\pi \tau_d^{QS}), 
\label{f_t_QS2}
\end{equation}
which is one of the main reasons why $f_T^{QS}$ is interpreted as a relevant FoM on how fast a digital FET works. In summary, the quasi-static approximation is valid whenever the condition $\Delta Q_1 \approx \Delta Q_2$ is satisfied. From (\ref{charge}), such condition can be equivalently written as $\Delta Q_3 \approx 0$. From (\ref{Q}), the previous conditions in a transient evolution means that the source current rapidly becomes equivalent to its high value $I_3(t) \approx I_3(\tau_d)$ while the drain current remains low $I_2(t) \approx I_2(0)$ during the intrinsic delay time interval $0<t<\tau_d$. These conditions are typical in many FETs with a large channel length $L$ where the intrinsic delay time $\tau_d$ is much larger than the temporal width of the current pulse generated by one electron $\tau_p$, i.e. $\tau_d>\tau_p$. Then, the total (particle and displacement) current in the drain and source contacts are detected only when electrons cross the surfaces $S_2$ and $S_3$, respectively. However, in FET devices with a short channel length $L$, one can easily get scenarios with  $\tau_d \approx \tau_p$ where an electron moving along the channel generates a time-dependent electric field that is detected as displacement current on the source and drain contacts without even crossing the surfaces $S_2$ (drain contact) and $S_3$ (source contact).  

\subsection{The Zero-order Non-quasi-static Definition of $f_T^{NQS}$}
\label{Zero}

In order to better include the drain displacement current, it seems more appropriate to use the same zero-order approximation of the exponential term,  $ e^{-j\omega t}\approx 1$,  that we have used for $Y_{11}$ in (\ref{Y_{11}1}), in the computation of $Y_{21}$ from (\ref{Yij})
\begin{align}
Y_{21}^{NQS}(\omega)& \approx g_m+\frac{j\omega}{\Delta V_1}\int_{0}^{\tau_d} (I_2(t)-I_2(\tau_d))\;{\rm d}t  \nonumber\\
&=g_m-j\omega |\Delta Q_2|/\Delta V_1 \label{Y_{21}2},
\end{align}
where $\Delta Q_2$ is also defined in (\ref{Q}). Consequently, from (\ref{current_gain1}), a non-quasi-static estimation ($NQS$) of $f_T$ gives \cite{Boots2004}
\begin{equation}
\label{f_T_st}
f_T^{NQS}=\frac{g_m}{2\pi \sqrt{\Delta Q_1^2-\Delta Q_2^2}/\Delta V_1}.
\end{equation}
This is a first step (zero-order Taylor approximation) in the evaluation of $f_T^{NQS}$ beyond the quasi-static approximation. In a typical n-type FET, when $\Delta V_1$ is positive, we can expect a positive transient current $I_2(t)$ satisfying $I_2(0) \leq I_2(t) \leq I_2(\tau_d)$,  while the current on the source is negative and decreases $I_3(0)\geq I_3(t) \geq I_3(\tau_d)$ because of the signs selected in Fig \ref{figure1}(b).  Since we deal with an increment of electrons (negative charge) in the channel, we expect $I_1(t) \geq 0$ in the metal. From (\ref{Q}) we get positive $\Delta Q_1$ and $\Delta Q_3$, while negative $\Delta Q_2$. Therefore, the expression $\Delta Q_1+\Delta Q_3=|\Delta Q_2|$ is achieved, which means  $|\Delta Q_1|<|\Delta Q_2|$. This condition will be numerically tested later. Therefore, the definition of $f_T^{NQS}$ in (\ref{f_T_st}) can be ill-defined because it deals with a square root of a negative number, that is, the condition $|h_{21}|=1$ cannot be reached with this zero-order non-quasi-static approximation.  

We arrive now to a relevant question about the adequacy of $f_T$ as a proper FoM for testing FET speed. \emph{Is it possible to find FETs where the gate phasor current is always smaller than the drain one, even with the exact definition of the $Y$ parameters in (\ref{current_gain1})? This would imply that, contrarily to what is assumed in the own definition of $f_T$, the current gain never drops to 0 dB at any frequency.} 

\section{Numerical simulation}
\label{simulation}

The conditions of validity of $f_T^{QS}$ were discussed in Section \ref{qs}. In Section \ref{Zero} we pointed out the possibility that the own definition of $f_T$ is ill-defined because there is no guarantee that the gate phasor current becomes higher than the drain phasor current as frequency grows. Next, we provide numerical confirmation of these drawbacks for ballistic nanoscale FETs.

\subsection{Device Structure and Time-dependent Simulations}

We will consider dual-gate FETs schematically drawn in Fig. \ref{figure1}(a) with a 2D channel material. These 2D materials are expected to improve electron mobility and to suppress the short channel effect for ultra-scaled devices. In order to simplify the numerical simulations (avoiding extra complications, like Klein tunneling or hole transport, that will obscure the interpretations of our numerical results), we will consider only electron transport in the conduction band of a n-type graphene-like material with a linear energy band $E_{\vec{k}}=\pm \hbar v_f |\vec{k}|$ being $v_f=5 \times 10^5$ m/s the Fermi velocity and $\vec{k}$ the wave vector which contains the two degrees of freedom $\{k_x,k_z \}$. The permittivity is $\epsilon=4\epsilon_0$ in the 2D material and $\epsilon=3.9\epsilon_0$ in dielectrics with $\epsilon_0$ is the vacuum permittivity. Electron transport will be assumed ballistic (without phonon or impurity scattering) and only the electron-electron interaction through the Poisson equation will be considered. The simulation box will not include the 3D-2D contact resistances and other parasitic elements (which are the well-known frequency bottleneck \cite{Schwierz2010}). Thus, we only simulate the intrinsic performance of FETs.

We will consider FET devices with a width of the current pulse associated to one electron comparable to the intrinsic delay time along the channel, i.e. $\tau_{p} \approx \tau_d$. These conditions just mean that the channel is short enough and the dielectric relaxation time large enough so that the displacement current of an electron crossing the channel has to be considered in each terminal even when the electron is in the middle of the channel. The extension $L'$ depicted in Fig. \ref{figure1}(a) is present to ensure the proper computation of such displacement current, even when the electrons are outside of the volume $\Omega$. The time-dependent total  currents in equation (\ref{total}) are computed with the BITLLES simulator \cite{BITLLES} from self-consistent Monte Carlo solutions of the Boltzmann and Poisson equations. The temporal step of the simulations is $\Delta t=7 \times 10^{-16}$ s. Finally, we notice that all the transient simulations have been repeated many times and the results properly averaged in order to minimize the presence of physical noise \cite{Zhan2016} (random fluctuations) in the current values. The reasons are to avoid extra non-pertinent complexities in the discussions of the results and to approach experimental $S$-parameters setups which provide measurements through several periods of the oscillating signals.

\subsection{Example 1: Device A}

\begin{figure}[t!]
\centering
\includegraphics[width=0.6\columnwidth]{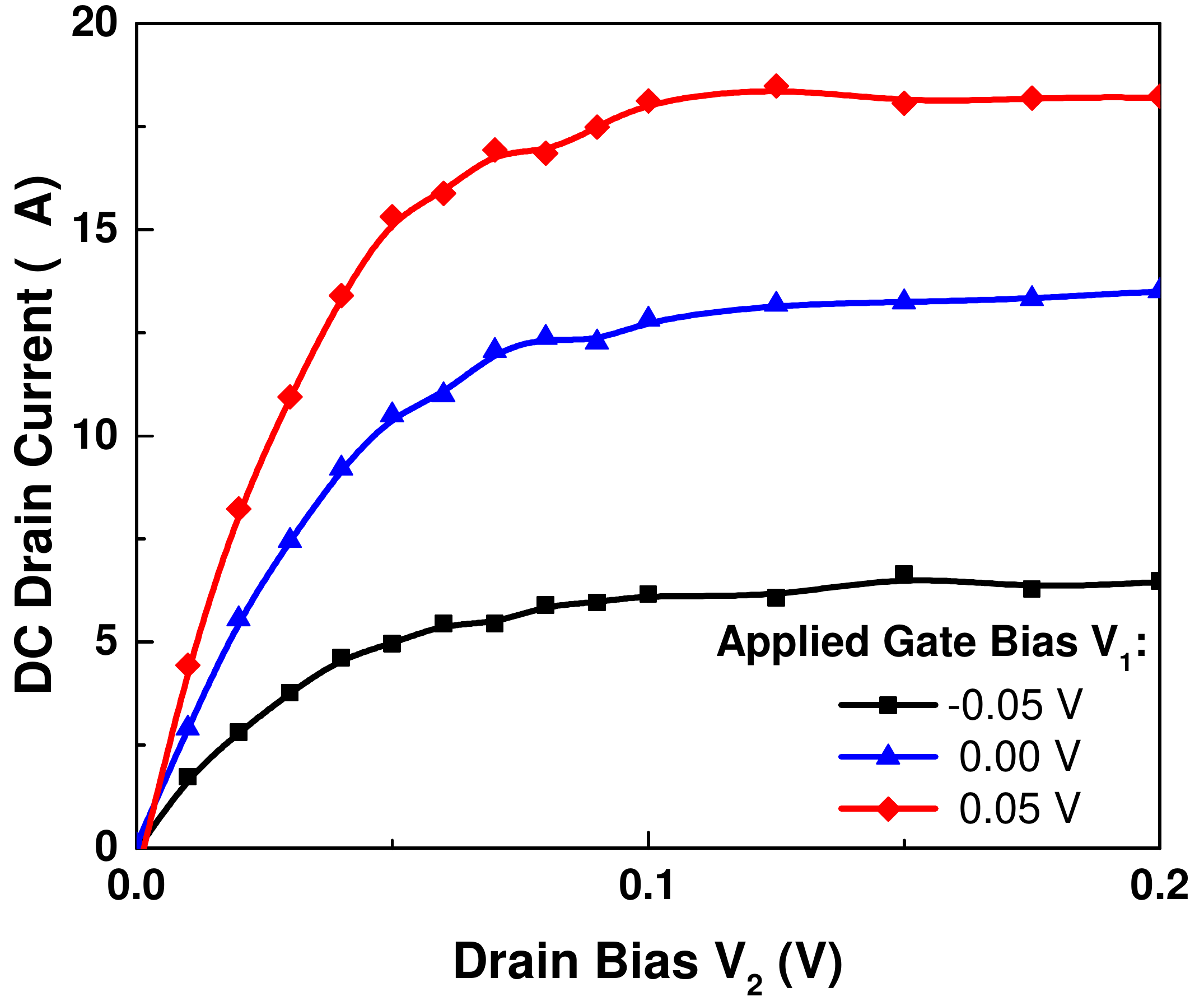}
\caption{The output characteristic of device A for different $V_1$. }
\label{figIVa}
\end{figure}
\begin{figure}[t!]
\centering
\includegraphics[width=0.65\columnwidth]{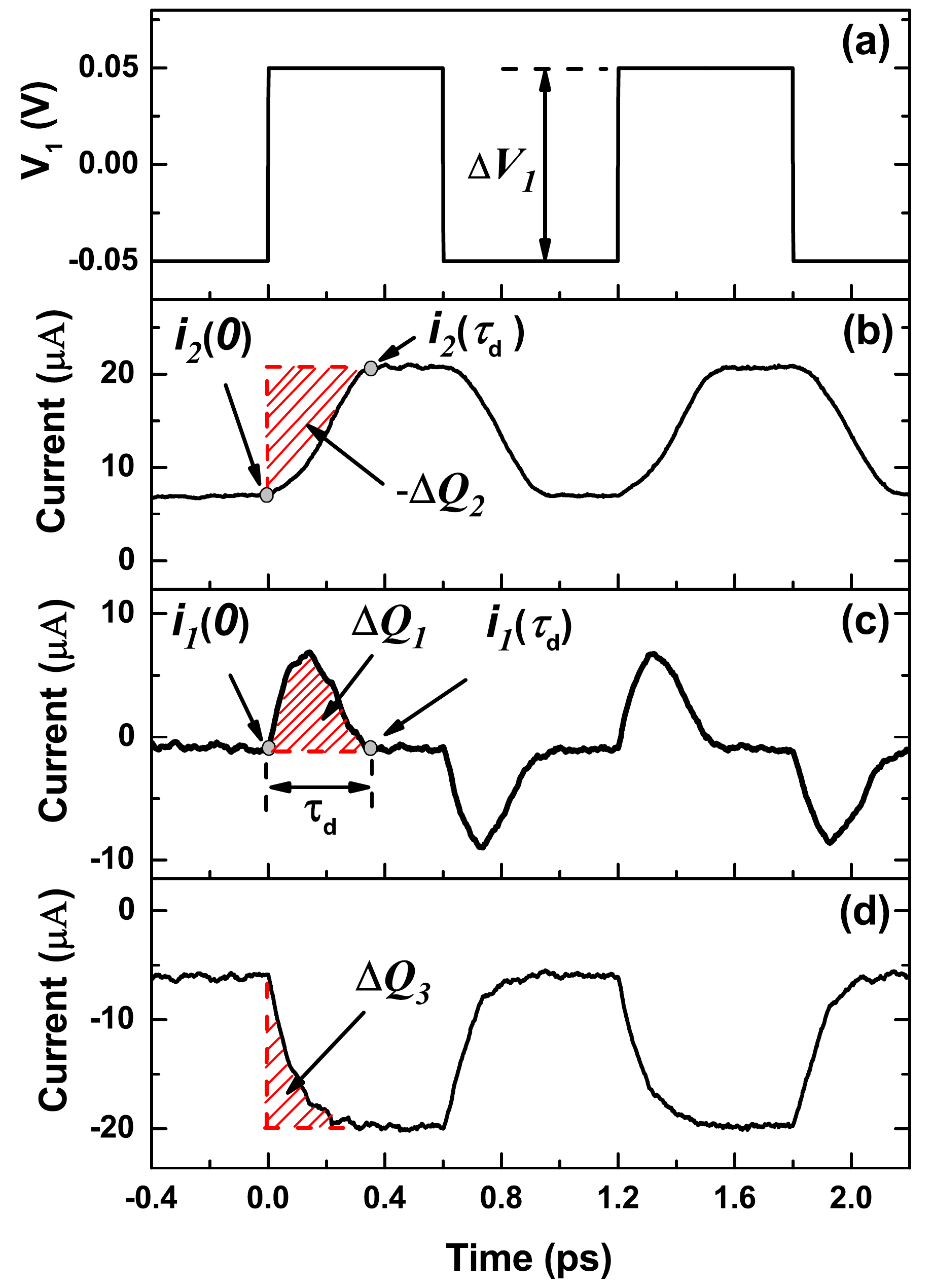}
\caption{Total (particle plus displacement) transient currents on the drain (b), gate (c) and source (d) contacts of  device A when a sequence of square voltage pulses (a) is applied on the gate contacts. $V_2=0.1$ V.}
\label{figure2}
\end{figure}
\begin{figure}[t!]
\centering
\includegraphics[width=0.60\columnwidth]{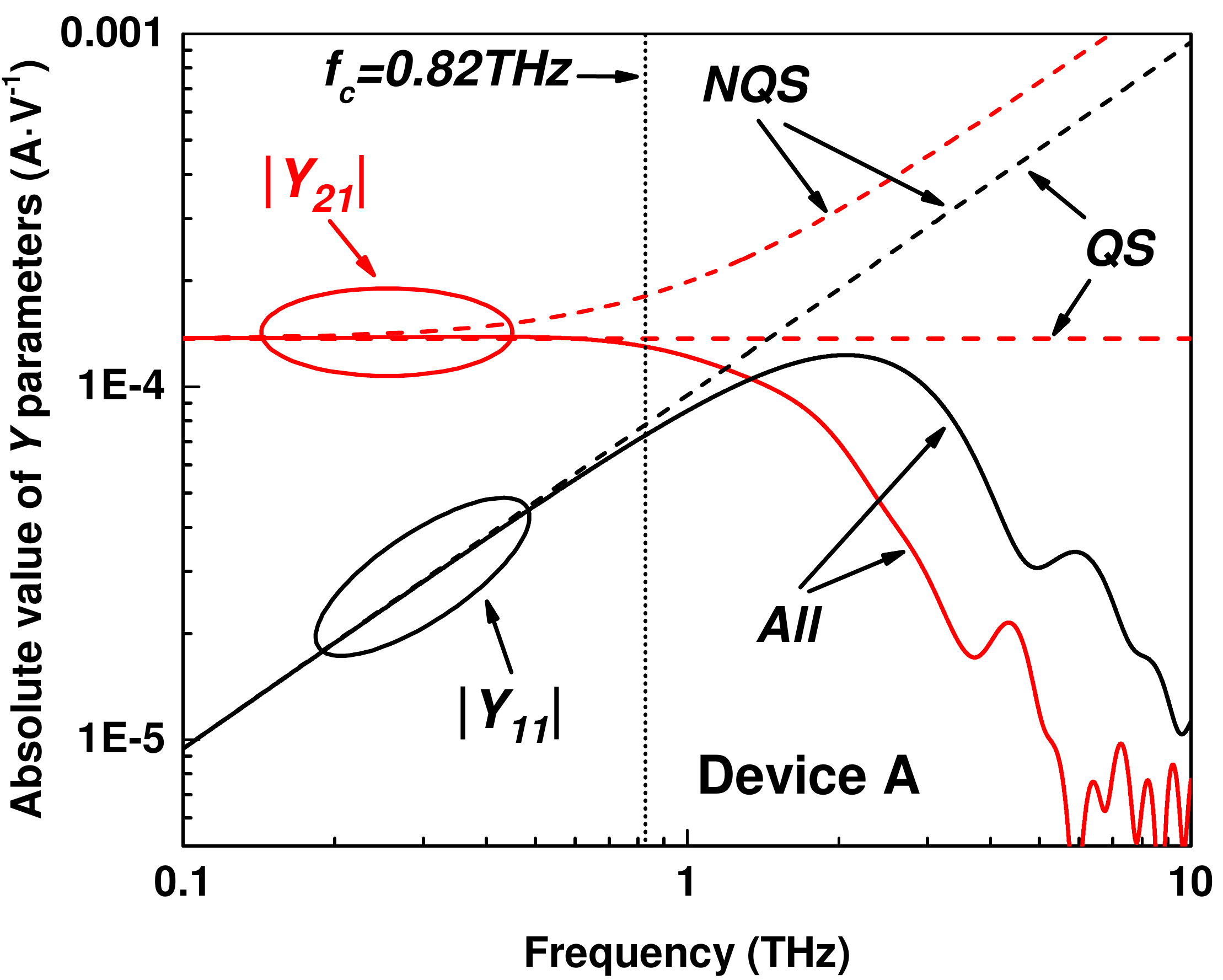}
\caption{The $Y$ parameters computed from the time-dependent simulation of the total (particle plus displacement) currents in device A using three different expressions. In solid lines, taken into account all orders (${All}$) and in dashed lines with the non-quasi-static ($NQS$) or the quasi-static ($QS$) approximations. }
\label{figure3}
\end{figure}

We consider device A with a volume $\Omega_A=100 \times (45 +1 +45)  \times 1125 \;\mathsf{nm}^3$ in Fig. \ref{figure1}(a) with the 2D material thickness $H=1$ nm and the gate length $L=100$ nm. In the simulation box, we set spatial steps $\Delta x=10$ nm, $\Delta y=11.25$ nm and $\Delta z=225$ nm resulting $22 \times 11 \times 7$ cells. The DC characteristic plotted in Fig. \ref{figIVa} is computed by time-averaging the total drain current in (\ref{total}) and by summing the net number of electrons transmitted through the drain surface. Both DC values coincide because the time-averaged displacement current is zero. Such double computation of the drain DC value certifies the correct simulation of the displacement current.     

The transient currents in response to two square voltage pulses on the gate contact are indicated in Fig. \ref{figure2}. As illustrated in Fig. \ref{figure1}(b), since we deal with a small-signal formalism, the evaluation of $Y_{21}$ and $Y_{11}$ is done with a DC bias, $V_2=0.1$ V, applied between drain and source contacts, and a DC voltage $V_1=-0.05$ V plus the transient perturbation $\Delta V_1=0.1$ V on the gate. In Fig. \ref{figure3}, the solid lines are $|Y_{21}^{All}|$ and $|Y_{11}^{All}|$ computed exactly from (\ref{Yij}) as a function of frequency. For frequencies higher enough, the absolute value of $Y_{21}^{All}$ shows strong frequency dependency and the $|Y_{11}^{All}|$ is no longer linearly increasing with the frequency, which is qualitatively identical to the experimental observations \cite{Cheng2001}. The values of $|Y_{21}^{All}|$ and $|Y_{11}^{All}|$ become equal at  $f_T=1.31$ THz.

Using the quasi-static approximation in expression (\ref{f_t_QS}), we get  $f_{T}^{QS}=1.45$ THz, which is similar to the previous value $f_T=1.31$ THz. In the non-quasi-static approximation, the formula (\ref{f_T_st}) requires the restriction $|\Delta Q_1| >|\Delta Q_2|$. As illustrated in Fig. \ref{figure2} (red dashed area), $\Delta Q_1=11.62 \times 10^{-19}$ C, $\Delta Q_2=-26.53 \times 10^{-19}$ C and $\Delta Q_3=14.91 \times 10^{-19}$ C. The result $\Delta Q_1<|\Delta Q_2|$ is coincident with what we anticipated in Section \ref{Zero}. Since $ \Delta Q_1^2<\Delta Q_2^2$, there is no solution to the $f_T$ in non-quasi-static case. This result can also be understood from (\ref{Y_{21}2}) indicating that $|\Delta Q_2|$ controls the frequency slope of $Y_{21}^{NQS}$ at high enough frequencies. Similarly, from (\ref{Y_{11}1}), the slope of $Y_{11}^{QS}$ is controlled by the $\Delta Q_1$. Since $\Delta Q_1<|\Delta Q_2|$, the terms  $Y_{21}^{NQS}$ and $Y_{11}^{QS}$ never cross as can be seen in Fig. \ref{figure3}. Surprisingly, the (zero-order) non-quasi-static model is even worse than the simpler quasi-static one.   

The errors when neglecting the displacement current in the computation of $f_T^{QS}$ can be quantified from expression (\ref{Y_{21}2}). The elimination of the drain displacement current can be justified for those frequencies satisfying that $g_m$ is larger or equal than the $\omega |\Delta Q_2|/\Delta V_1$. By imposing the previous condition, $g_m \approx 2 \pi f_c |\Delta Q_2|/\Delta V_1$, we get a definition of the maximum frequency $f_c$ where the drain displacement current can be reasonably neglected
\begin{equation}
 f_c=\frac{g_m}{2\pi |\Delta Q_2|/\Delta V_1}.
\label{fc}
\end{equation}
 However, since we have demonstrated in equation (\ref{charge}) that $|\Delta Q_2|>\Delta Q_1$, we always get $f_c < f_T^{QS}$ which can be seen in Fig. \ref{figure3} where solid lines (\textit{All}) start to deviate from dashed lines (\textit{QS}) at $f_c=0.82$ THz $< f_{T}^{QS}=1.45$ THz. Notice that the condition $\Delta Q_1 \approx \Delta Q_2$ in (\ref{fc}) implies that $f_c \approx f_T^{QS}$ justifying the arguments on the range of validity of $f_T^{QS}$ mentioned in Section \ref{qs}.

\subsection{Example 2: Device B}

\begin{figure}[t!]
\centering
\includegraphics[width=0.6\columnwidth]{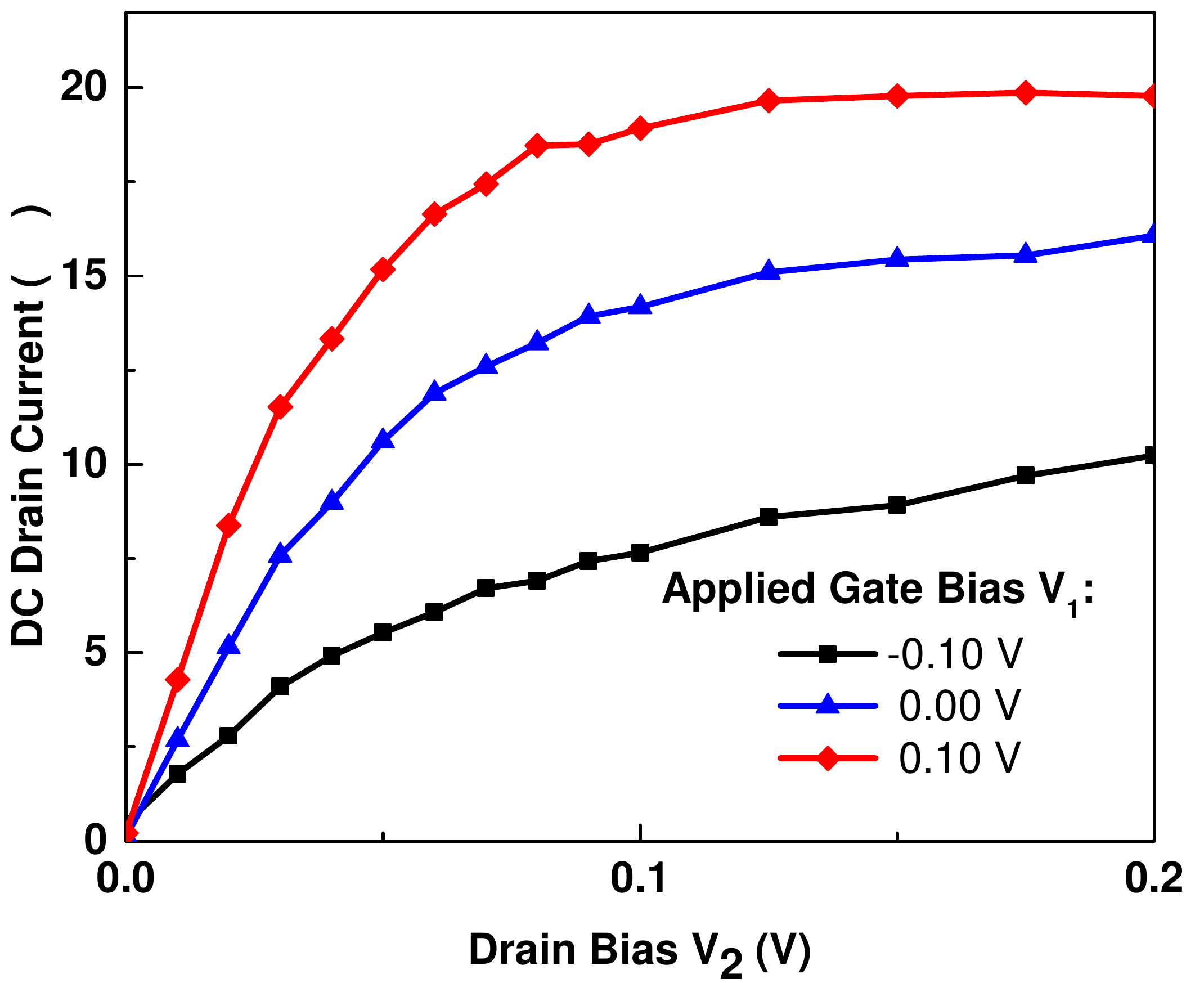}
\caption{The current-voltage characteristic of device B for different $V_1$. }
\label{figIVb}
\end{figure}
\begin{figure}[t!]
\centering
\includegraphics[width=0.65\columnwidth]{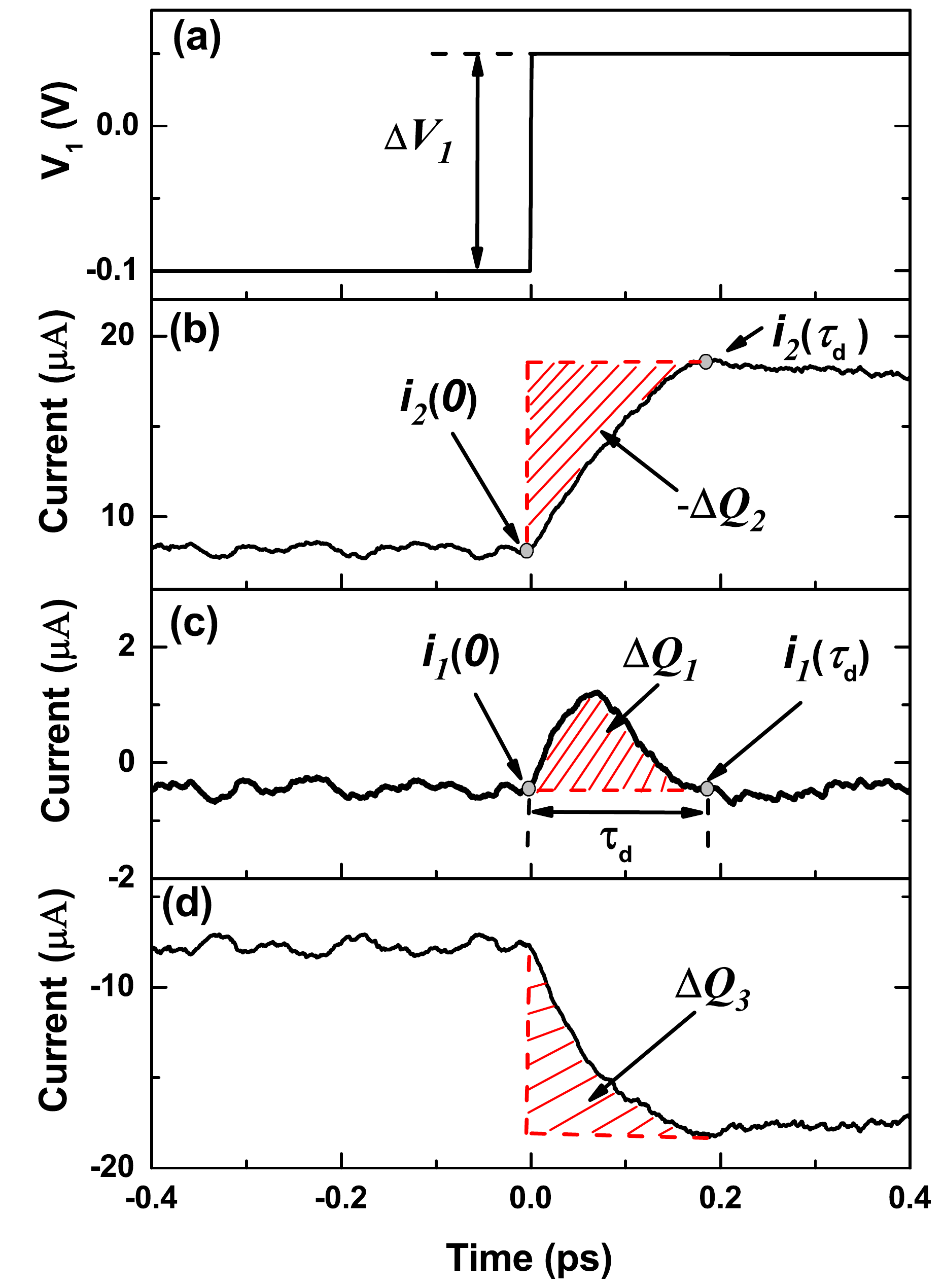}
\caption{The total (particle plus displacement) transient currents on the drain (b), gate (c) and source (d) contacts of  device B in response to a step voltage perturbation (a) on the gate contacts. $V_2=0.1$ V.} 
\label{figure7}
\end{figure}
\begin{figure}[t!]
\centering
\includegraphics[width=0.60\columnwidth]{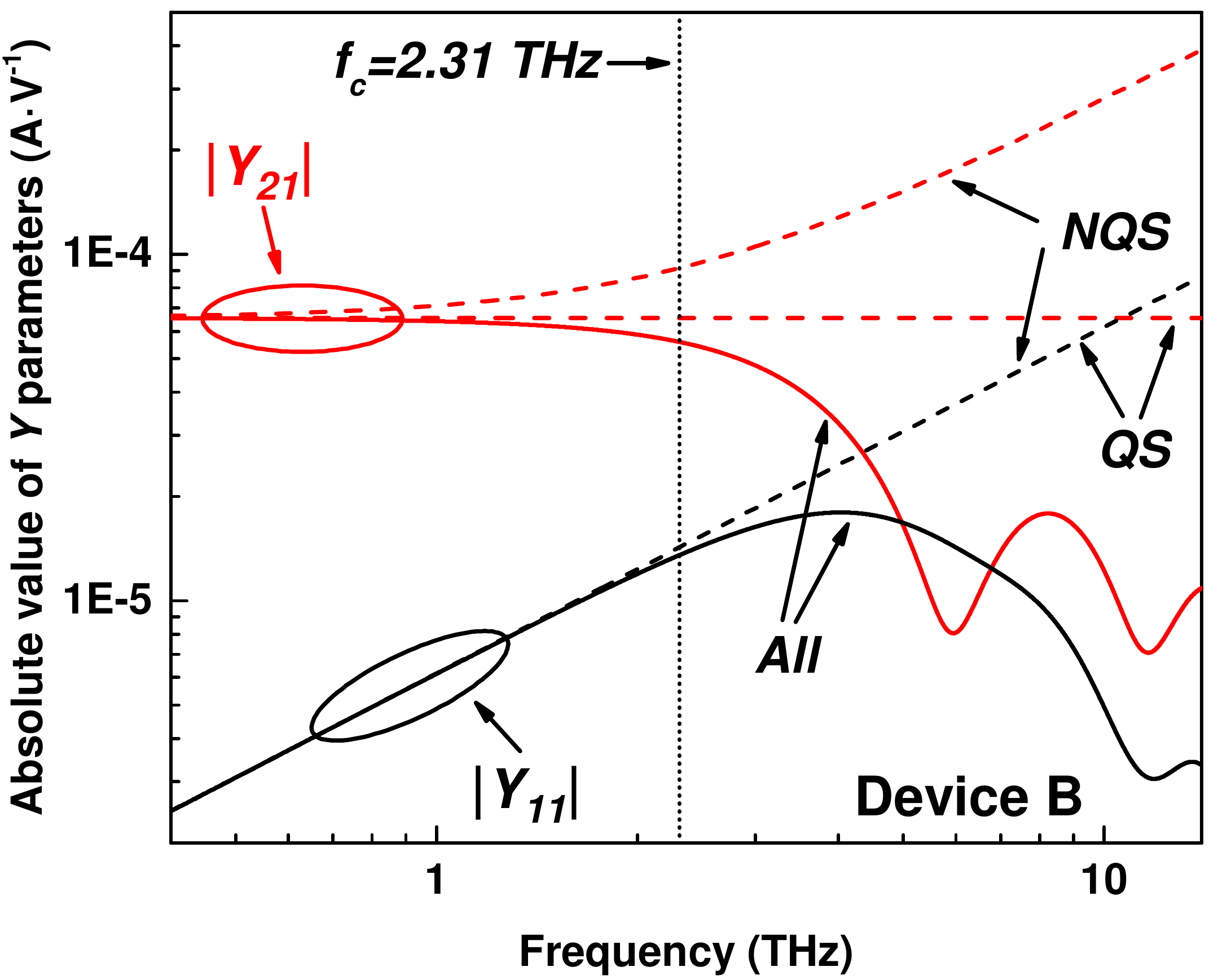}
\caption{The $Y$ parameters computed from time-dependent simulation of the total currents in device B using three different expressions (as in Fig. \ref{figure3}).} 
\label{figure4}
\end{figure}

The quasi-static approximation seems to imply that a desired condition for a fast FET is $\Delta Q_1 \rightarrow 0$ ($C_1\rightarrow 0$) if short channel effects are still under control. Such condition would imply that $f_{T}^{QS} ={g_m}/({2\pi \Delta Q_1/\Delta V_1})={g_m}/({2\pi C_1}) \rightarrow \infty$ in equation (\ref{f_t_QS}) and $\tau_d^{QS} \approx C_1 \Delta V_1 / \Delta I_2 \approx \Delta Q_1/\Delta I_2 \rightarrow0$. We consider a new design (device B) with the goal of getting $C_1\rightarrow 0$. In particular we consider the same FET of Fig. \ref{figure1}(a) with the geometry $\Omega_B=20\times (45+1+45) \times 700 \;\mathsf{nm}^3$ (gate length $L=20$ nm) under the same type of simulation as in device A. In the simulation box, we set spatial steps $\Delta x=2$ nm, $\Delta y=11.25$ nm and $\Delta z=140$ nm resulting $22 \times 11 \times 7$ cells. We plot the DC current-voltage characteristic of device B in Fig \ref{figIVb}. In spite of the small capacitance, the short channel effects are reasonably under control. We use $V_2=0.1$ V, applied between drain and source contacts, and $V_1=-0.1$ V plus the transient perturbation $\Delta V_1=0.15 $ V on the gate. The total transient currents on the drain, gate and source contacts of device B due to a step voltage perturbation in the gate are plotted in Fig. \ref{figure7}, where the sum of the total currents also maintains consistency with the continuity equation (\ref{transient_current}). Moreover, the incremental charge $\Delta Q_1=1.54 \times 10^{-19}$ C, $\Delta Q_2=-6.76 \times 10^{-19}$ C and $\Delta Q_3=5.24 \times 10^{-19}$ C. 

The $Y$ parameters changing with frequency in all orders (solid lines), NQS and QS models (dashed lines) for device B are plotted in Fig. \ref{figure4}.  In the quasi-static estimation, because the $|\Delta Q_1|$ in device B becomes smaller than that of device A, the $ Y_{11}^{QS}$ is shifted towards the horizontal axis, therefore, the quasi-static value of the cut-off frequency increases giving $f_{T}^{QS}=10.64$ THz. The behaviors of NQS for device B and device A are similar. In the all orders model, the condition $|h_{21}|=1$ is satisfied at $f_T=4.97$ THz. The oscillations of $Y_{11}^{All}$ and $Y_{21}^{All}$ at higher frequencies are the Fourier transform of time-dependent variations of the total currents in Fig. \ref{figure7}, which can be associated to plasmonic oscillations with shorter periods than the value of $\tau_d$ plotted there. So the exact value of the $f_T$ is randomly influenced by such oscillations. We have assumed an ideal metallic contact (dielectric relaxation time equals to zero) in all simulations. One can expect significantly different randomness in the oscillations of the $Y$ parameters at high frequency for heavily doped contacts \cite{Oriols2016}. The problem present in device B is that the real frequency where the device stops working properly is much lower than $f_T^{QS}=10.64$ THz and $f_T=4.97$ THz.

\subsection{Example 3: Device C}

\begin{figure}[t!]
\centering
\includegraphics[width=0.60\columnwidth]{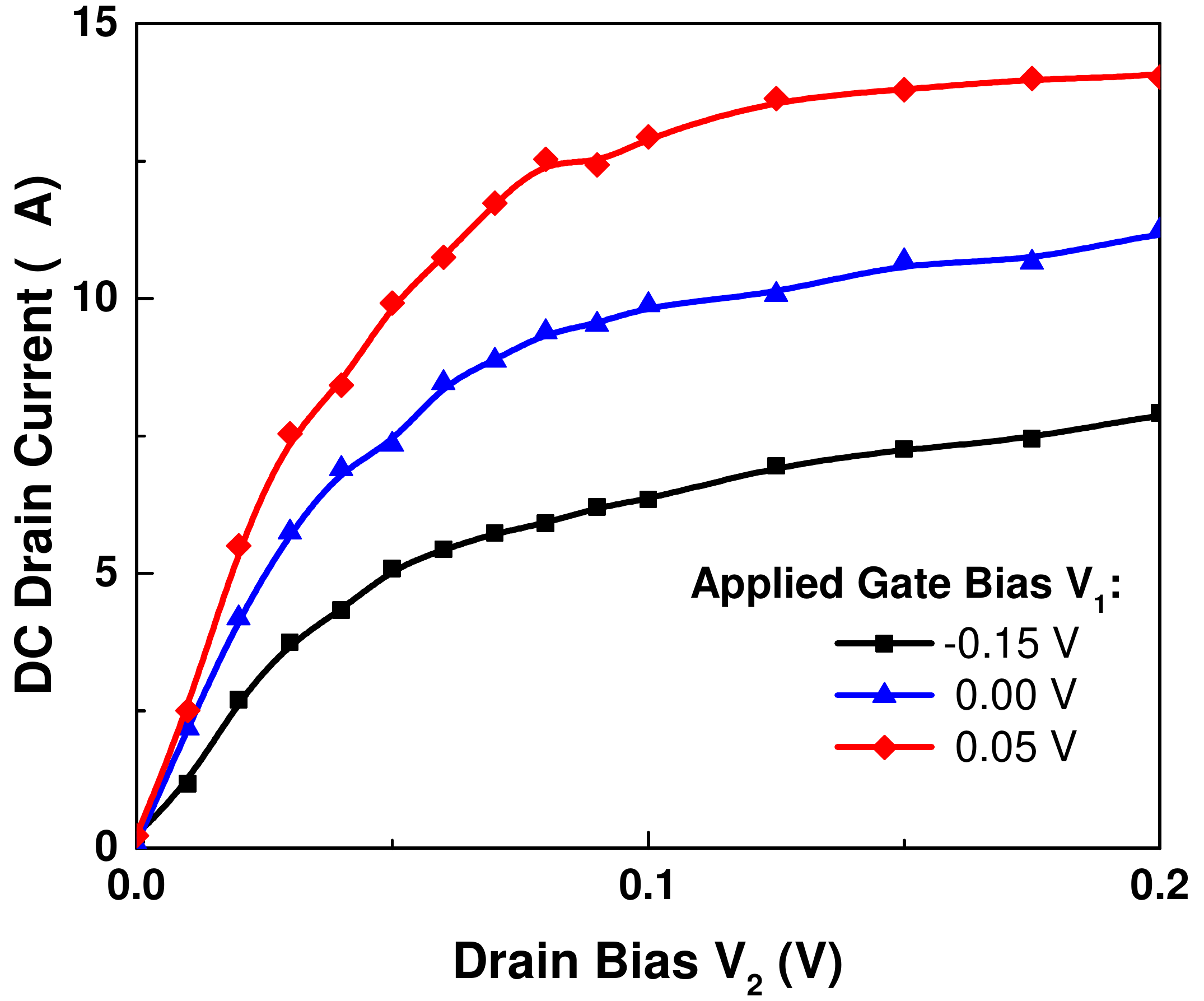}
\caption{The current-voltage characteristic of device C for different $V_1$. }
\label{figIVc}
\end{figure}

\begin{figure}[t]
\centering
\includegraphics[width=0.65\columnwidth]{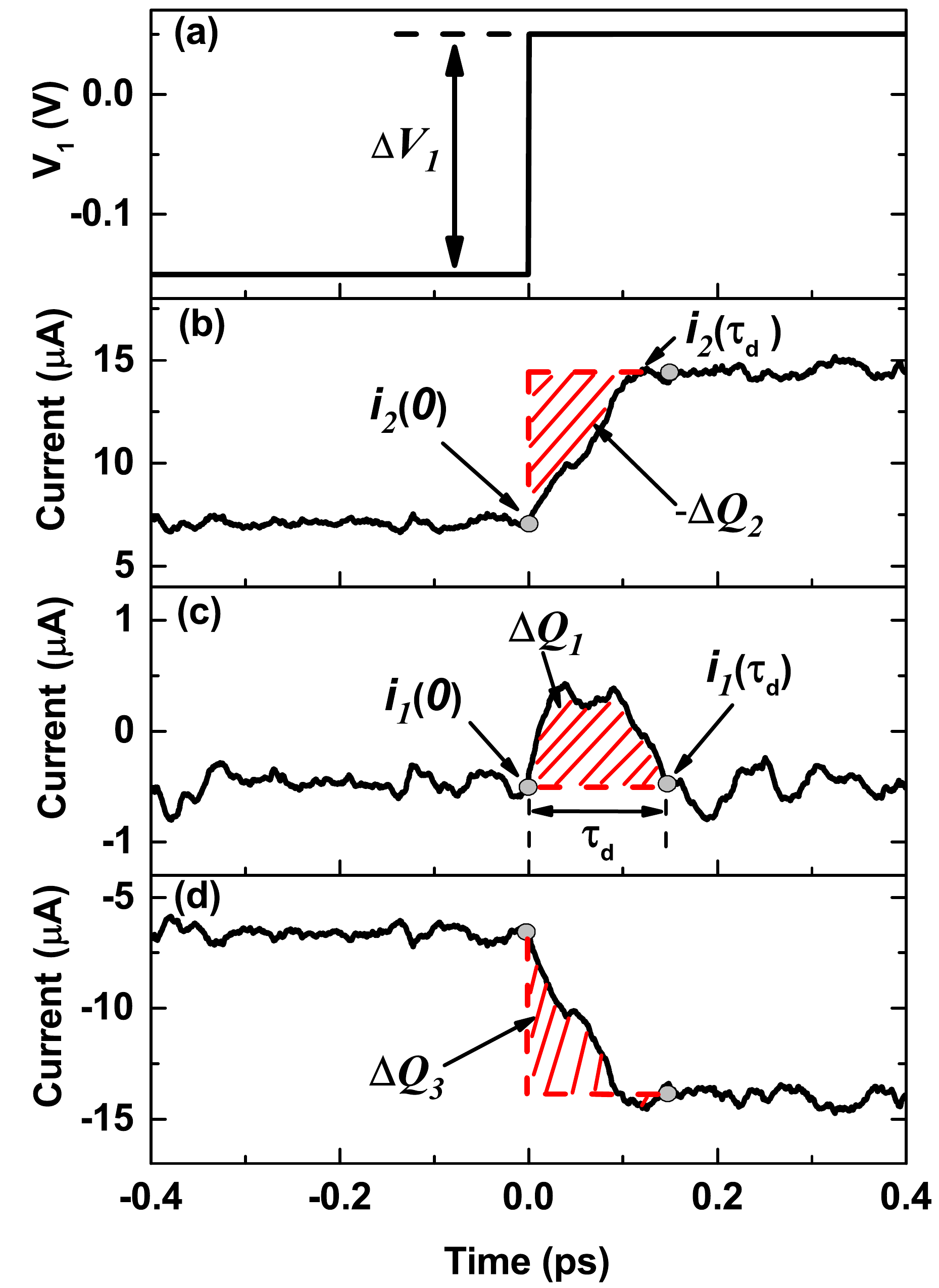}
\caption{The total (particle plus displacement) transient currents on the drain (b), gate (c) and source (d) contacts of  device C in response to a step voltage perturbation (a) on the gate contacts. $V_2=0.1$ V.} 
\label{figureC}
\end{figure}

\begin{figure}[t]
\centering
\includegraphics[width=0.60\columnwidth]{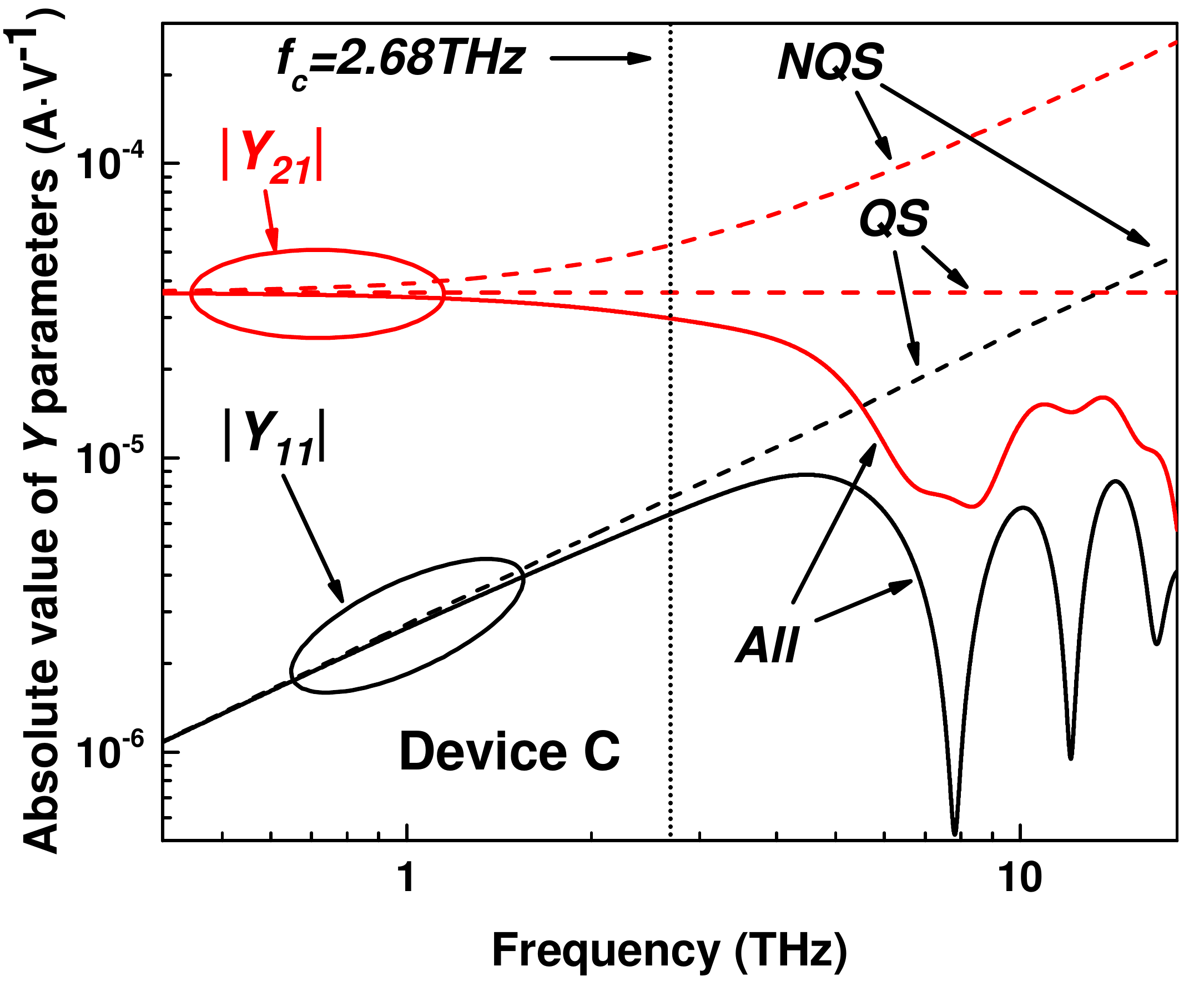}
\caption{The $Y$ parameters computed from time-dependent simulation of the total currents in device C using three different expressions (as in Fig. \ref{figure3}).}
\label{figure10}
\end{figure}

To go a step further, we design a device C which has volume $\Omega_C=20\times (45+1+45) \times 200 \;\mathsf{nm}^3$ (gate length $L=20$ nm). The DC current is plotted in Fig. \ref{figIVc}, which indicates the ability of the gate to control the channel.  In the transient current plotted in Fig. \ref{figureC} in response to a step voltage perturbation, we get the incremental charge $\Delta Q_2=-4.33 \times 10^{-19}$ C, $\Delta Q_3=3.45 \times 10^{-19}$ C and $\Delta Q_1=0.9 \times 10^{-19}$ C. The $\Delta Q_1$ approaches to 0. The $Y$ parameters are plotted in Fig. \ref{figure10}. The QS result $f_T^{QS}=13.43$ THz is quite far from the maximum frequency $f_c=2.68$ THz. It is relevant to emphasize that the condition $|h_{21}|=1$ is never satisfied in the whole THz window for the all orders model, i.e. the $|Y_{21}^{All}|$ and $|Y_{11}^{All}|$ never cross, showing a misleading result $f_T=\infty$. Certainly, the real nano FET stops working properly at some frequency. 

To understand the real speed limitation of devices B and C, let us notice that the simulations in Figs. \ref{figure2}, \ref{figure7} and \ref{figureC} can be interpreted in two different ways. First, as we have done up to here, the currents are the FET response to a step voltage perturbation needed for computing the small-signal $Y$-parameter in a linear system.  Second, they can be identically interpreted as the current response when the gate change from the digital voltage `0' to `1'. Then, the time $\tau_d$ defined in expression (\ref{Q}) is directly related to the intrinsic delay time $\tau_d$ discussed in the Introduction as a FoM for digital electronics. In other words, the two pulses in Figs. \ref{figure2}(a) can be understood as the input digital signal and the drain currents as the output digital signals. Certainly, the FETs are not properly switched-off in our small-signal simulations, and a large-signal simulation will be needed, in principle. However, the present simulations are enough to compare the different FET speed estimators.  The relation between the input and output signals in Figs. \ref{figure2}, \ref{figure7} and \ref{figureC} can be modeled from linear system theory. It is clear that the relevant frequencies are inversely proportional to the time interval $\tau_d$. A reasonable expression could be $f^{d}_T \approx 1/(2\tau_d)$ \cite{Carlson} \cite{Tsividis}. 

In device A, from Fig.  \ref{figure2}, we get a value of the intrinsic delay time $\tau_d=0.352$ ps, resulting $f^{d}_T=1.42$ THz, which is similar to the quasi-static cut-off frequencies $f^{QS}_T=1.45$ THz. However, in device B, from Fig. \ref{figure7}, we get a simulated $\tau_d=0.185$ ps giving $f^{d}_T=2.70$ THz, which is much smaller that the value $f_{T}^{QS}=10.64$ THz. In device C, the result is even worse where the intrinsic delay time is $\tau_d=0.138$ ps providing $f_T^d=3.62$ THz. The reason why the quasi-static expression (\ref{f_t_QS}) does not capture the real intrinsic delay time in devices B and C is because the approximation $\Delta Q_2 \approx \Delta Q_1$ is no longer true. In fact, since we looked for $\Delta Q_1 \rightarrow 0$, the result $\Delta Q_2 >> \Delta Q_1$ gives $f_T^{QS}$ in (\ref{f_t_QS}) much larger than $f_c$ in (\ref{fc}). Let us notice that the quasi-static estimation of the intrinsic delay time is again clearly misleading. In device B, we get $\tau_d^{QS} \approx C_1 \Delta V_1 / \Delta I_2 \approx 0.015$ ps, which is more than one order of magnitude smaller than our simulated value $\tau_d=0.185$ ps (the ratio $\tau_d/\tau_d^{QS}$ is different from $f_T^d/f_T^{QS}$ because of the factor $\pi$ in (\ref{f_t_QS2})). In device C, the $\tau_d^{QS}\approx 0.012$ ps is also more than one order of magnitude shorter than the simulated value $\tau_d=0.138$ ps. The reason of this discrepancy is also the condition $\Delta Q_2 >> \Delta Q_1$, which invalidates the quasi-static estimation discussed in Section \ref{qs}

\section{Final discussion and conclusions}

In conclusion, we have established the condition for the validity of the quasi-static approximation of $f_T^{QS}$ in terms of the electrical current and electrical flux on the gate, drain and source FET terminals defined in expressions (\ref{total}) and (\ref{Q}). Such approximation is applicable when $\Delta Q_1 \approx \Delta Q_2$ which means that we are dealing with FETs where the intrinsic delay time is much larger than the temporal width of the current pulse generated by one electron, i.e. $\tau_d>\tau_p$ (with large channel length $L$ as in device A). On the contrary, in devices where $\tau_d \approx \tau_p$ (i.e. as in devices B and C with short channel length $L$), the quasi-static approximation is not applicable because the electric field generated by electrons are not screened inside the device active region, and its associated displacement current becomes relevant during all the time while the electron is traversing the channel. We have shown through analytical arguments supported by numerical simulations that the estimations of the intrinsic cut-off frequency based on  $|h_{21}|=1$ (with the quasi-static $f_T^{QS}$, zero-order non-quasi-static $f_T^{NQS}$ or without approximations $f_T$) can provide misleading results for the speed of FETs. This problem is specially severe for nanoscale FETs which are routinely modeled from quantum transport simulators. The explicit quantum simulation of the time-dependent displacement current and $\tau_d$ demand such huge amount of computational resources \cite{Oriols2007,Oriols2013,Oriols2016} that the intrinsic FoM of the speed of such ballistic FET are routinely taken from quasi-static estimations. As shown in some examples in this work, such type of quasi-static estimations can erroneously predict the FET speed by one order of magnitude. Other examples show no finite value of $f_T$ underlying an important limitation of the traditional definition of $f_T$ to properly quantify the speed of FETs. However, when parasitic elements are included in the simulation, one can expect a tendency to recover the validity of the quasi-static approximation ($\tau_d$ grows and $\tau_p$ remains the same) at the price of getting lower FET speed than its intrinsic value.

\ifCLASSOPTIONcaptionsoff
  \newpage
\fi

\end{document}